\documentstyle[aps,prl,preprint]{revtex}

\begin{document}
\draft
\title{A different view of the quantum Hall plateau-to-plateau 
transitions}
\author{D. Shahar, D. C. Tsui and M. Shayegan}
\address{Department of Electrical Engineering, Princeton University,
Princeton New Jersey, 08544}
\author{E. Shimshoni}
\address{Department of Mathematics-Physics, Oranim-Haifa 
University, Tivon 36006, Israel}
\author{S. L. Sondhi}
\address{Department of Physics, Princeton University,
Princeton New Jersey, 08544}

\date{\today}

\maketitle
\begin{abstract}  
We demonstrate experimentally that the transitions between adjacent 
integer quantum Hall (QH) states are equivalent to a QH-to-insulator 
transition occurring in the top Landau level, in the presence of
an inert background 
of the other completely filled Landau levels, each contributing 
a single unit of quantum conductance, $e^{2}/h$, to the total Hall 
conductance of the system.
\end{abstract}


One of the most active topic of research in the field of 
two dimensional electron 
systems has been the study of the transition regions 
separating adjacent quantum Hall (QH) states
\cite{wei,engel,Koch91,Engel:MW1,HPWei:Current,Sondhi:RMP}.
These transition 
regions were considered mainly in the theoretical framework of 
scaling\cite{Sondhi:RMP,Huckestein:rev}, extending to the 
high magnetic field ($B$) QH
regime a methodology that was originally developed 
for the study of the $B=0$ metal-insulator transition\cite{Gang4}.

A comprehensive experimental study by Wei and his collaborators
\cite{wei}
revealed much of the fascinating physics of these transitions. By 
focusing on the diagonal resistivity ($\rho_{xx}$) peaks separating 
adjacent QH minima, and on the Hall resistivity ($\rho_{xy}$) steps 
accompanying them, they were able to show that in the transition 
regions, the conduction process can be well-described by that 
expected of a system near a critical point of a quantum phase 
transition. Two of their experimental findings particularly
support a quantum-critical description of 
these transitions; both were obtained from a 
study of the evolution of the width of the $\rho_{xx}$ peaks, 
$\Delta B$, and the maximal slope of the $\rho_{xy}$ steps, 
$(\frac{d\rho_{xy}}{dB})^{max}$, as a function of temperature, $T$.
Their first result was that the width and inverse-slope approach zero as 
$T\rightarrow 0$,
\begin{equation}
\Delta B,(\frac{d\rho_{xy}}{dB})^{-1}_{max}\sim T^{\kappa},
\label{peakScaling}
\end{equation}
suggestive of proximity to a scale-invariant critical point at a 
critical field $B=B_{c}$ and $T=0$.
Their second observation was that the value of $\kappa$ is 
independent of which Landau level (LL) is involved (for spin split LL's),
and was also the same, 
within error, for many samples studied. This {\em universality} is another 
expected signature of a quantum phase transition\cite{Sondhi:RMP}.

More recently, another class of transitions in the quantum Hall regime 
were studied\cite{Jiang93,Alphenaar:2terminalHI,Shahar:Univ,Wong95}.
These $B$-driven transitions are not between adjacent QH states. 
Rather, they are transitions from QH states to the insulating 
phase that terminates the QH series. Since, in the insulator, 
$\rho_{xx}\rightarrow \infty$, these transitions are not characterized 
by a $\rho_{xx}$ peak, and experimentally they appear, on first sight, 
to be different from the inter-QH, plateau-to-plateau transitions. 
It was soon realized, however, that a critical $B$ exists for 
these transitions as well and that universal scaling behavior 
is also observed 
in their vicinity, characterized by a critical exponent that is
consistent with the $\kappa$ obtained for the inter-QH transitions
\cite{Wong95,Shahar:PUthesis}. 
For these QH-to-insulator transitions, there is 
significant experimental evidence 
in support of the existence of the theoretically 
expected\cite{YHuo93} universal 
critical amplitude, with the $\rho_{xx}$ value at the transition point 
close to $h/e^{2}$\cite{Shahar:Univ,Wong95}. This should be 
contrasted with the inter-QH transitions,
for which most researchers report a critical amplitude that is not only 
significantly ($40-80\% $) smaller than the theoretically 
expected value, but in 
many cases is also $T$-dependent\cite{Rokhinson:peaks}. 

From a theoretical standpoint, the transitions to the insulator are 
similar to the inter-QH transitions; both occur as the fermi 
energy crosses the extended states that exist in the center of a 
LL. They differ, however, in that the inter-QH transitions 
take place in the presence of a number of filled LLs 
underneath the fermi energy, separated by a gap from the top LL where the 
action takes place. If one assumes that 
the only contribution 
of the background LLs is to the Hall conductance, each filled LL 
adding $e^{2}/h$ to the measured value of $\sigma_{xy}$, it may be 
possible, by numerically removing that contribution from the 
experimental data, to test this equivalence. There is an equivalent 
scenario involving  fractional QH states, 
in which the transitions take the form of a QH-to-insulator
transition of a set of quasiparticles in parallel with a background
(parent) QH liquid. This insight has been systematized 
theoretically\cite{KLZ}, and the general idea of relating the 
transitions in the integer and fractional regimes has been corroborated 
by recent experiments\cite{Shahar:Univ,Wong95}. 
In this paper we will only be concerned with integer QH transitions.

The purpose of this communication is to report on an experimental 
test of the conjecture that the inter-QH and the QH-to-insulator
transitions are indeed similar. For simplicity, we focus on 
the transition from the $\nu=2$ to $\nu=1$ integer QH states
(dubbed $2-1$), which 
occurs at the top spin-split first LL,
and compare it to the $\nu=1$-to-insulator ($1-0$) 
transition in the same sample, which takes place near the center of 
the lower spin-split first LL. To implement this comparison, we
utilized a straightforward scheme, as follows. First, we 
obtain traces of $\rho_{xx}$ and $\rho_{xy}$ for the $2-1$ 
transition using a Hall-bar shaped sample etched in a high-density 
($n=2.27\cdot 10^{11}$ cm$^{-2}$), low-mobility 
($\mu=10.8 \cdot 10^{3}$ cm$^2$/Vsec), MBE grown GaAs/AlGaAs wafer.
These resistivity traces, taken at several $T$'s,
are plotted in Fig. 1a. The transition is typified by 
a $\rho_{xx}$ peak that widens with $T$ and by the accompanying step 
in $\rho_{xy}$. Next,
we convert the $\rho$'s to $\sigma$'s using the 
standard matrix conversion,
\begin{equation}\label{Sig}
\sigma_{xx(xy)}=\frac{\rho_{xx(yx)}}{\rho_{xx}^{2}+\rho_{xy}^{2}}\,
\end{equation}
and plot the $\sigma$ traces in Fig. 1b. 
We then obtain the conductivity of the topmost LL by subtracting
from the conductivity data the contribution of the lowest, full LL:
\begin{equation}
\sigma_{xx}^{t}=\sigma_{xx}
\end{equation}

\begin{equation}
\sigma_{xy}^{t}=\sigma_{xy}-e^{2}/h\,
\end{equation}
assuming, as mentioned, that the only 
contribution of the lowest LL is $e^{2}/h$ to the Hall conductivity 
(Throughout this paper, the 
index $t$ refers to the contribution of the topmost LL to 
the transport coefficients). Next we 
convert $\sigma_{xx}^{t}$ and $\sigma_{xy}^{t}$ to new 
resistivities, $\rho_{xx}^{t}$ and $\rho_{xy}^{t}$, which are the 
resistivities of the topmost LL. This completes 
our procedure, and we are now ready for the comparison with the data 
obtained from the $1-0$ transition in the same sample.

The comparison is made in Fig. 2, where we plot
$\rho_{xx}^{t}$ (solid lines) and $\rho_{xy}^{t}$ (short-dashed lines)
as a function of $\nu$ for the $2-1$ transition (Fig. 2a), and
traces of $\rho_{xx}$ and $\rho_{xy}$ vs. $\nu$ obtained from 
the same sample near the $1-0$ 
transition terminating the QH series, in Fig 2b (here, of course, 
$\rho^{t}=\rho$). 
While for the $\rho_{xx}$ traces in both graphs of Fig. 2 we present 
data at our lowest $T$ range ($T<150$ mK), 
the $\rho_{xy}$ traces shown were taken at an elevated $T$ ($\approx 
320$ mK) for which reliable data can be obtained. The difficulties 
with the Hall component data at lower $T$'s, shown in Fig. 3, 
will be discussed below.

The first point that can be observed in Fig. 2 is the clear similarity of 
the overall appearance of the traces in the two graphs. In particular, 
both sets of $\rho_{xx}$ traces are characterized by a $T$-independent 
crossing point
of the traces taken at different $T$'s which, for the $1-0$ transition,
has been identified as the QHE-to-insulator transition point.
\cite{Alphenaar:2terminalHI,Shahar:Univ} 
Building on the similarity of the traces obtained from the different 
transitions, it is natural to associate the $2-1$ transition $\nu$, 
$\nu_{c}$, with 
the crossing point of the $\rho_{xx}^{t}$ traces observed in Fig 2a.  
Adopting this identification of $\nu_{c}$, we proceed to explore its 
consequences on the $\rho$ and $\sigma$ traces of Fig. 1.
We wish to emphasize that the similarity also holds for the 
$\rho_{xy}$ traces, both remaining nearly constant through the 
transition region, and near their 
quantized value of $h/e^{2}$\cite{Shahar:dualitySC,Shahar:natureHI}. 
We also note that initial search for duality symmetry
\cite{Shahar:natureHI,Shahar:dualitySC,shimshoni} for the 
resistivities of the top LL showed that $\rho_{xx}^{t}$ indeed exhibit 
duality symmetry for the $2-1$ transition.

First, we note that the transition $\nu$ is {\em not} at the $\rho_{xx}$ 
peak (the dashed line in Fig. 1 marks $\nu_{c}$).
In fact, the position of the $\rho_{xx}$ peak is clearly $T$ 
dependent, at the same $T$ range where the crossing point in 
$\rho_{xx}^{t}$ (see Fig. 2b) is not. 
Instead, a $T$-independent point can be seen 
at the high-$\nu$ shoulder of the $\rho_{xx}$ peaks. It is this point, rather 
than the peak's center, that coincides with the transition as 
inferred from the $\rho_{xx}^{t}$ traces. On the other hand, 
inspecting the $\sigma_{xx}$ traces in Fig. 1b reveals that the 
$\sigma_{xx}$ peak is centered around $\nu_{c}$, and its position is 
much less $T$-dependent. One can immediately draw the following 
conclusions: a) associated with the $2-1$ transition is a 
$\sigma_{xx}$ peak, whose position is $T$-independent (at low $T$), and 
is centered around $\nu_{c}$, b) a peak in $\rho_{xx}$ 
also exist in the vicinity of the transition, but its center is 
offset by a $T$-dependent amount from $\nu_{c}$, and c) on the
$\rho_{xx}$ peak, the transition 
point is characterized by a $T$-independent point located at its 
shoulder. The fact that only one of either the $\rho_{xx}$ or the 
$\sigma_{xx}$peak is symmetric is hardly surprising 
because obtaining one from the other involves the distinctly 
asymmetric Hall component. Here, we demonstrate that at low 
$T$, the $\sigma_{xx}$ peak is the symmetric one. 

Inspecting Fig. 1, one can identify the existence of crossing 
points in the Hall components, $\sigma_{xy}$ and $\rho_{xy}$. For the 
$\sigma_{xy}$'s, the value at the crossing point is close to the 
mid-value between the two QHE plateaus\cite{YHuo93}, 
but for the $\rho_{xy}$'s it 
is not so, with the transition point clearly at much lower value. The 
$\nu$ position of these crossing points coincides with $\nu_{c}$, as it 
is identified from the crossing points of the $\rho_{xx}$ traces.
As we noted in previous 
works\cite{Shahar:natureHI,Shahar:dualitySC,shimshoni},
there appear to be a new relation 
between the diagonal and Hall transport coefficients near QH 
transitions. This relation implies that 
$\rho_{xy}^{t}$ remains constant and equal to its quantized value 
across the transition, into the insulating phase. Our experimental 
observations are in support of 
a constant $\rho_{xy}$ near both, the $1-0$ transition, and the 
$1/3$-fractional QH to insulator transition. The constancy of the 
$\rho_{xy}^{t}$ trace of Fig. 2a provides further evidence to the 
equivalence of the $1-0$ and $2-1$ transitions. An equivalent way of 
expressing this relation that easily lends itself to a more 
detailed experimental test, was put forth by Ruzin and his
co-workers\cite{ruzin}. 
Their results are cast in the form of a `semicircle' law 
for the conductivities:

\begin{equation}
(\sigma^{t}_{xx})^{2}+(\sigma^{t}_{xy})^{2}=\sigma^{t}_{xy}.
\label{semicircle}
\end{equation}
To test this we plot, in Fig. 3, 
$\sigma^{t}_{xy}$ vs. 
$(\sigma^{t}_{xx})^2+(\sigma^{t}_{xy})^2$for traces obtained 
at several $T$'s from $26-222$ mK. Overall, the data 
follow the expected straight line at higher $T$'s, with systematic 
deviations at low $\sigma_{xy}$ that becomes more significant as $T$ 
is lowered. At present, we can not unambiguously
delineate the source of these 
deviations\cite{Shahar:natureHI}; whether the semicircle law 
correctly describes the QH
transitions as $T\rightarrow 0$, or is just an approximation 
applicable at higher $T$'s, is outside the reach of the current 
experiment.

So far we discussed the similarity of the $2-1$ and $1-0$ transitions 
on a qualitative level. In order to extend our discussion in a more 
precise way, we now address the quantitative parameters describing the 
transitions, namely the critical exponents and amplitudes. In Fig. 4 
we present a scaling analysis of our $\rho_{xx}$ data for both 
transitions. Using a scaling procedure similar to that used by Wong et 
al.\cite{Wong95} we plot $\rho_{xx}$ for the $1-0$ transition 
(Fig. 3a) and $\rho_{xx}^{t}$ for the $2-1$ transition (Fig 3b) 
versus the scaling argument, $(\nu-\nu_{c})/T^{\kappa}$. While 
$\nu_{c}$ can be directly obtained from the data, we vary the value 
of $\kappa$ until we obtain the optimal ``collapse'' of the $\rho_{xx}$ 
traces obtained at different $T$'s. The resulting values of the critical 
exponent, $\kappa=0.45\pm 0.05$, are the same for both transitions 
and are also in good agreement with results obtained in previous 
studies of QHE transitions\cite{wei,Wong95}. 
It is worthwhile to re-emphasize the 
ubiquitous nature of $\kappa$, and its independence of the number of 
filled LL where the transition takes place.

Another quantitative aspect of the transitions is their critical 
resistivity. Both transitions depicted in Fig. 2 have a value of the 
critical resistivity at $B_{c}$ close to $h/e^{2}$, again in good 
agreement with studies of the $1-0$ transitions\cite{Shahar:Univ}.
As mentioned before, 
the issue of the universal critical resistivities for the inter-QHE is 
still under debate. One may argue that the process of eliminating 
the contribution of the lowest LL from the $2-1$ data may 
result in an incorrect critical resistivity value for the 
$\rho_{xx}^{t}$ obtained for that transition. To alleviate this concern 
we carefully inspect the raw $\rho_{xx}$ data presented in Fig. 1a. We 
recall that the scaling theory of the QH transitions predicts not 
one, but two, distinct universal critical 
amplitudes for the $2-1$ transition.  
The first is $\rho_{xx}^{c}$, the value of $\rho_{xx}$ 
at $B_{c}$, and the second is $\rho_{xx}^{p}$, the peak value of 
$\rho_{xx}$ which, as discussed above, is distinct from $\rho_{xx}^{c}$.
As can be seen, both are $T$-independent at the low-$T$ range shown, 
with $\rho_{xx}^{c}=4.4$ k$\Omega$ and $\rho_{xx}^{p}=6.25$ k$\Omega$.
Both these values are within 20\% of the 
expected\cite{fn1}
values of $\frac{h}{5e^{2}}$ and $\frac{h}{4e^{2}}$, 
respectively, lending further support to the universal 
character of the transitions.

To summarize, in this work we have demonstrated 
the equivalence of the inter-QH, 
plateau-to-plateau transitions, to the QH-to-insulator transition 
that terminates the QH series, and tested the proposed semicircle 
relation between the longitudinal and Hall components of the 
conductivity tensor near the transitions.
\noindent

This work was supported by the Israeli Academy of Sciences (ES), the
A. P. Sloan Foundation and NSF grant \# DMR-9632690 (SLS), and the NSF.


\begin{references}
\bibitem{fn1}In the critical region the resistivities are expected to be
universal functions of the combination $(B-B_c)/T^{1/\nu z}$.
Assuming that the non-universal amplitude multiplying this
combination is the same for both components, it follows that
$\rho_{xx}$ is a universal function of $\rho_{xy}$. The form
of this function is fixed if the semi-circle law is assumed
and can be seen to lead to the maximum value $\rho^p_{xx}={h 
\over 4e^2}$ \cite{ruzin}. Additionally, $\rho^c_{xx}={h \over
5e^2}$ follows from generalizing the lowest Landau level results
of Ref. \cite{YHuo93} to the second LL.
\end{references}

\begin{figure}
\caption{(a) $\rho_{xx}$ (lower curves) and $\rho_{xy}$ vs. $\nu$ 
taken in the vicinity of the $\nu=2$ to $1$ transition, at $T=42$, 
$70$, $101$ and $137$ mK. Note the narrowing of the transitions as $T$ 
is lowered.  (b) $\sigma_{xx}$ and $\sigma_{xy}$ vs. $\nu$, 
calculated from (a).Dashed line in both (a) and (b) 
indicates $\nu_{c}$, inferred from the data in Fig. 
2a (see text).}
\end{figure}

\begin{figure}
\caption{(a) $\rho^{t}_{xx}$ (solid lines) and $\rho^{t}_{xy}$ 
(long-dashed line) for the $2-1$ transition, calculated from the 
data in Fig. 1a. The $T$ for the $\rho^{t}_{xx}$ data
 are 42, 70, 101 and 137 mK, and for the $\rho^{t}_{xy}$ trace $T=330$ 
 mK. (b) Measured $\rho_{xx}$ (solid lines) and $\rho_{xy}$ (long-dashed line) 
 for the $1-0$ transition. The $T$ for the $\rho_{xx}$ data are 42, 
 84, 106 and 145 mK, and for the $\rho_{xy}$ trace $T=323$ mK. 
 Dashed line in both (a) and (b) indicates the transition $\nu$ inferred 
 from the common crossing point of the $\rho^{t}_{xx}$ (or 
 $\rho_{xx}$) traces.}
\end{figure}

\begin{figure}
\caption{A plot of $\sigma^{t}_{xy}$ vs. 
$(\sigma^{t}_{xx})^2+(\sigma^{t}_{xy})^2$ for the $2-1$ transition. A 
straight line indicates a semicircle relation between the conductivity 
components.
}
\end{figure}

\begin{figure}
\caption{A scaling analysis of the $\rho^{t}_{xx}$ (a) $\rho_{xx}$ 
(b) data in Figs. 2a and 2b, respectively. The scaling exponent 
$\kappa$, is determined to within 20\%.
}
\end{figure}

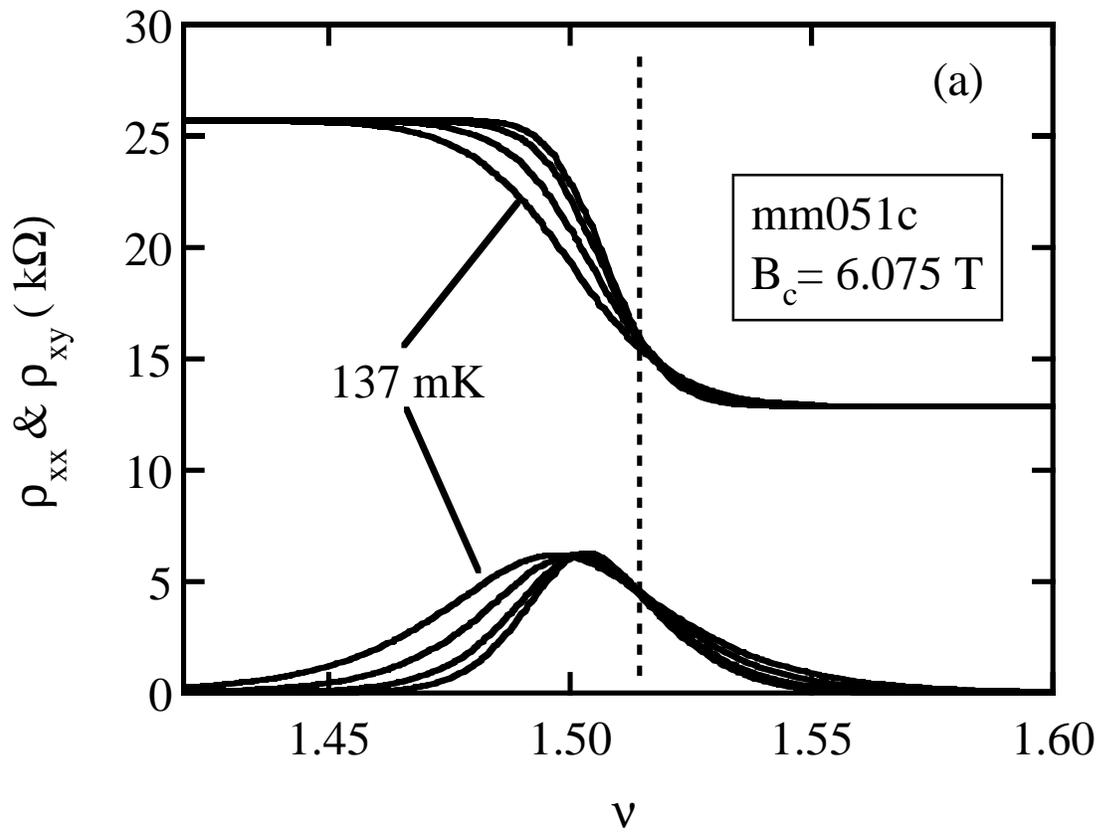
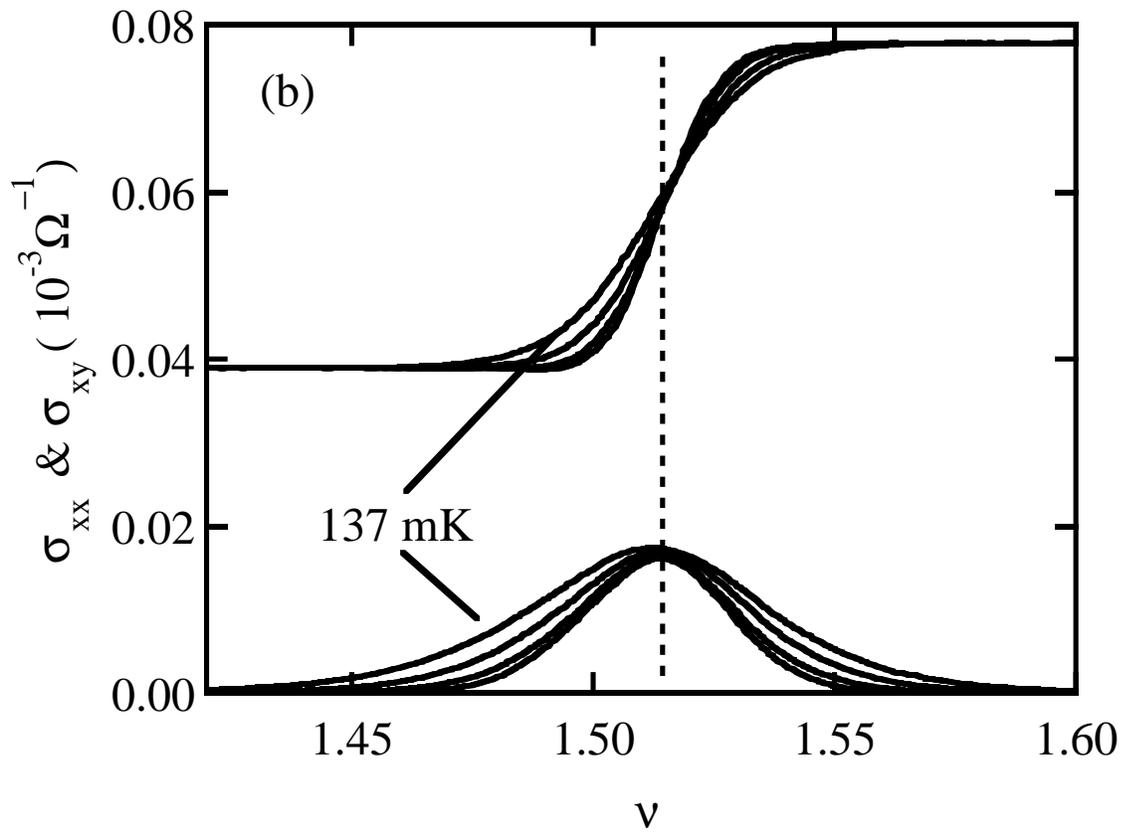
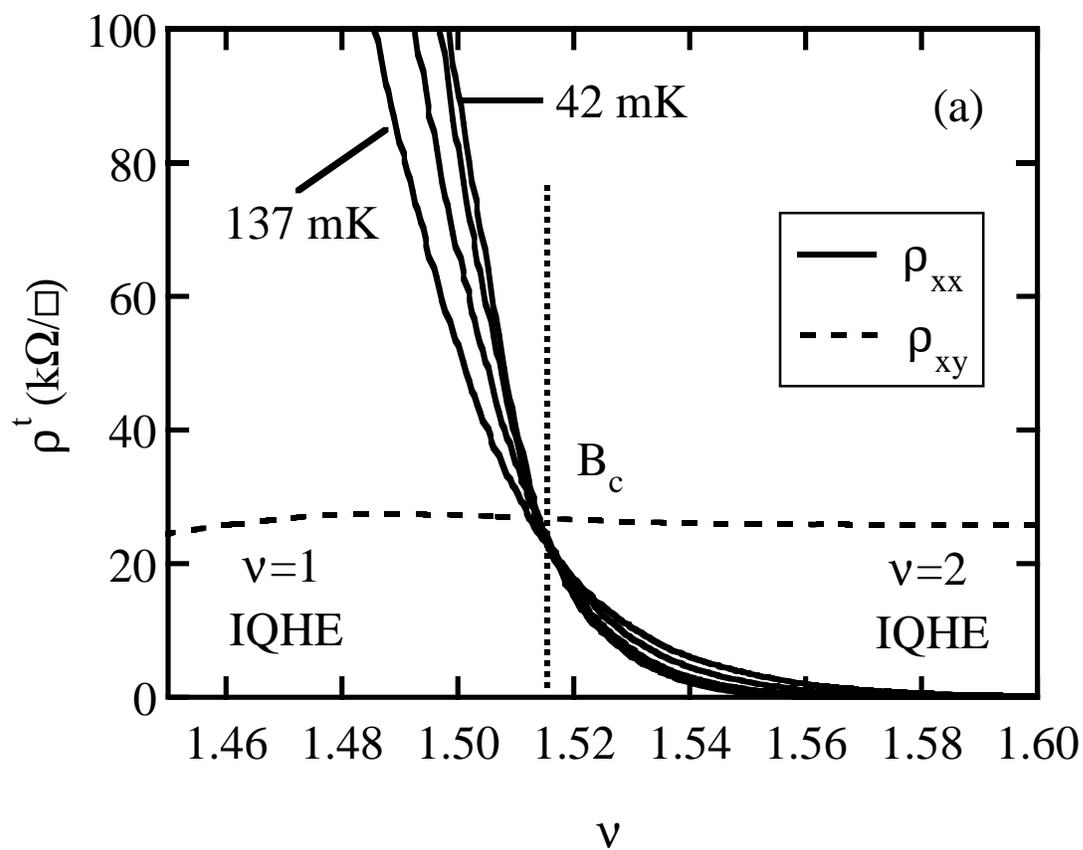
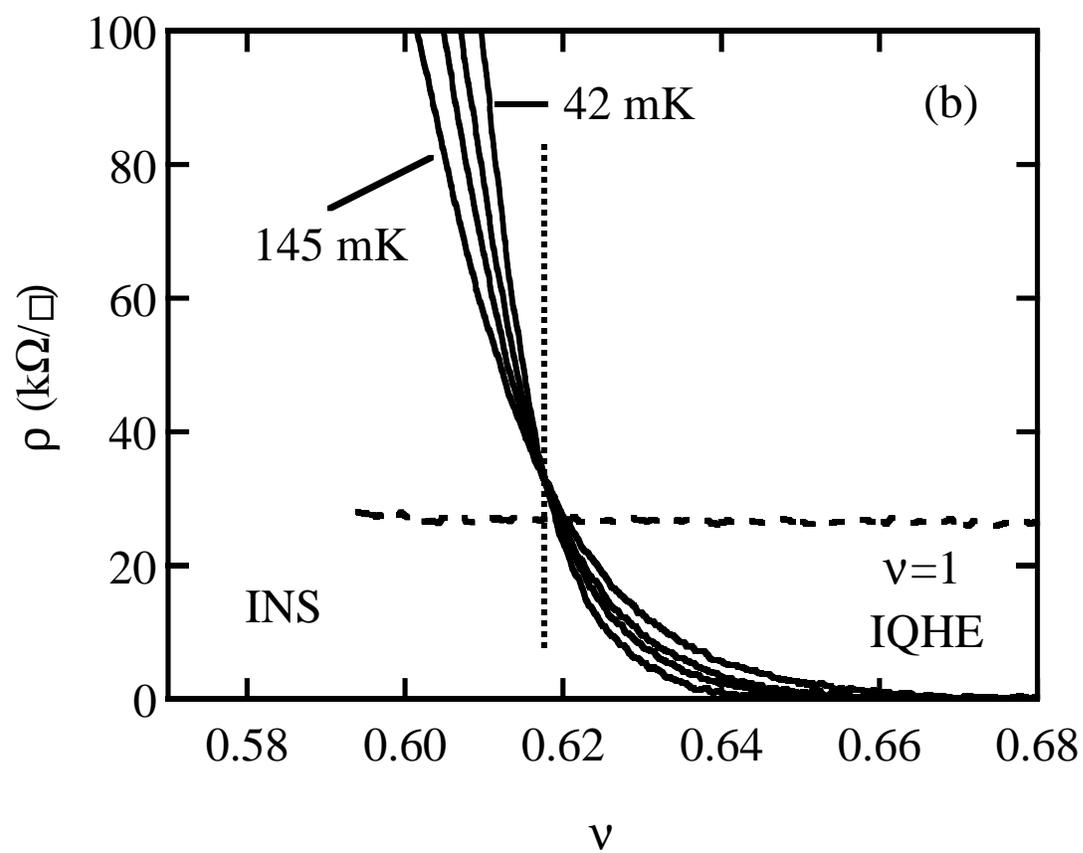
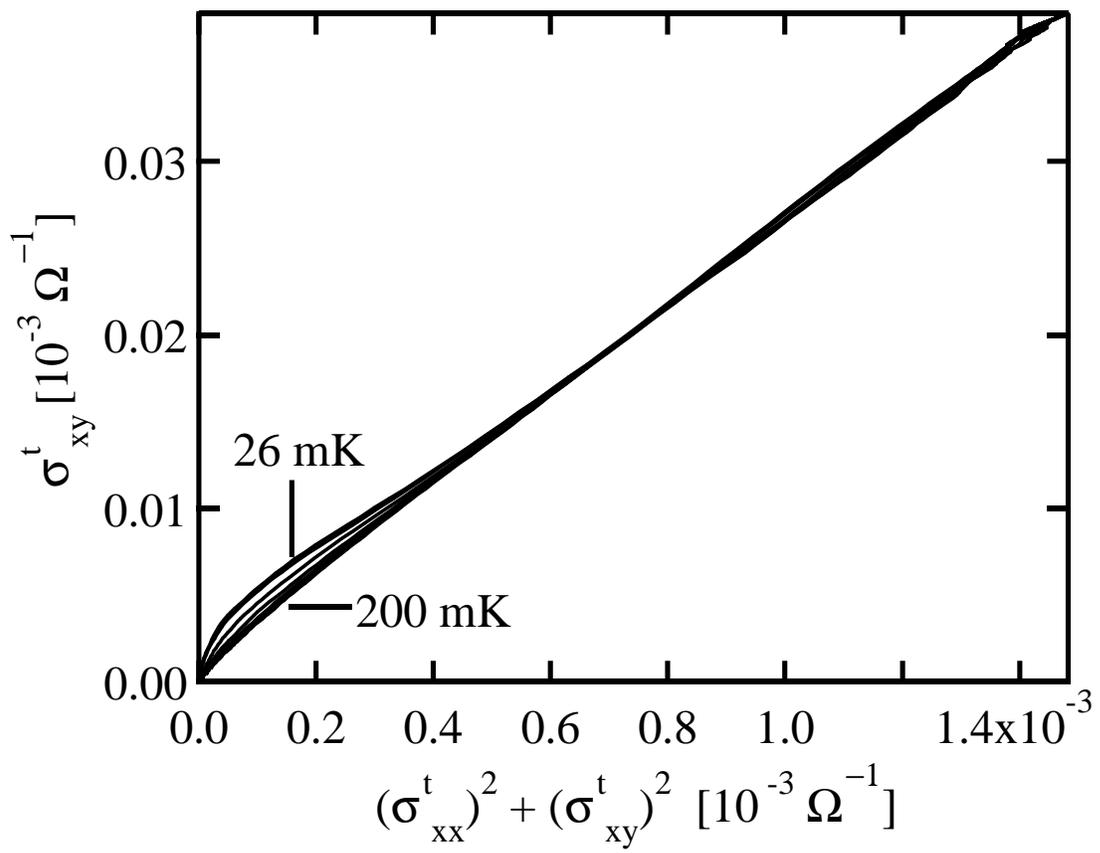
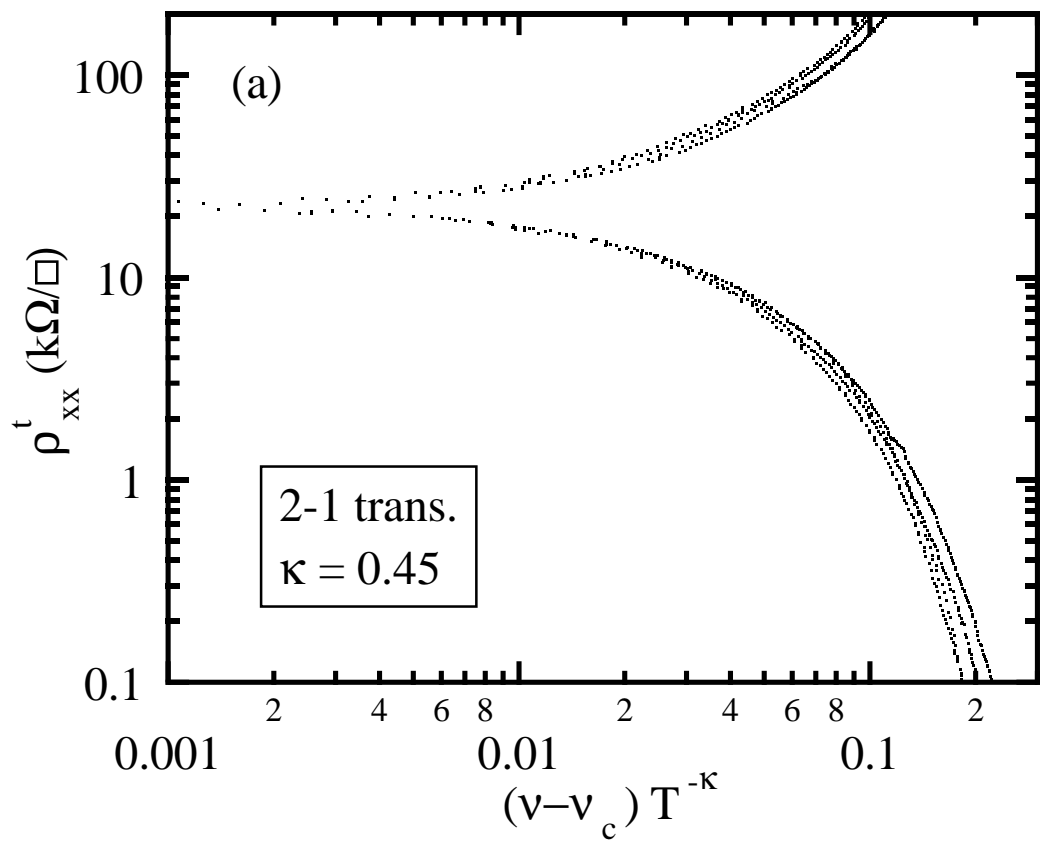
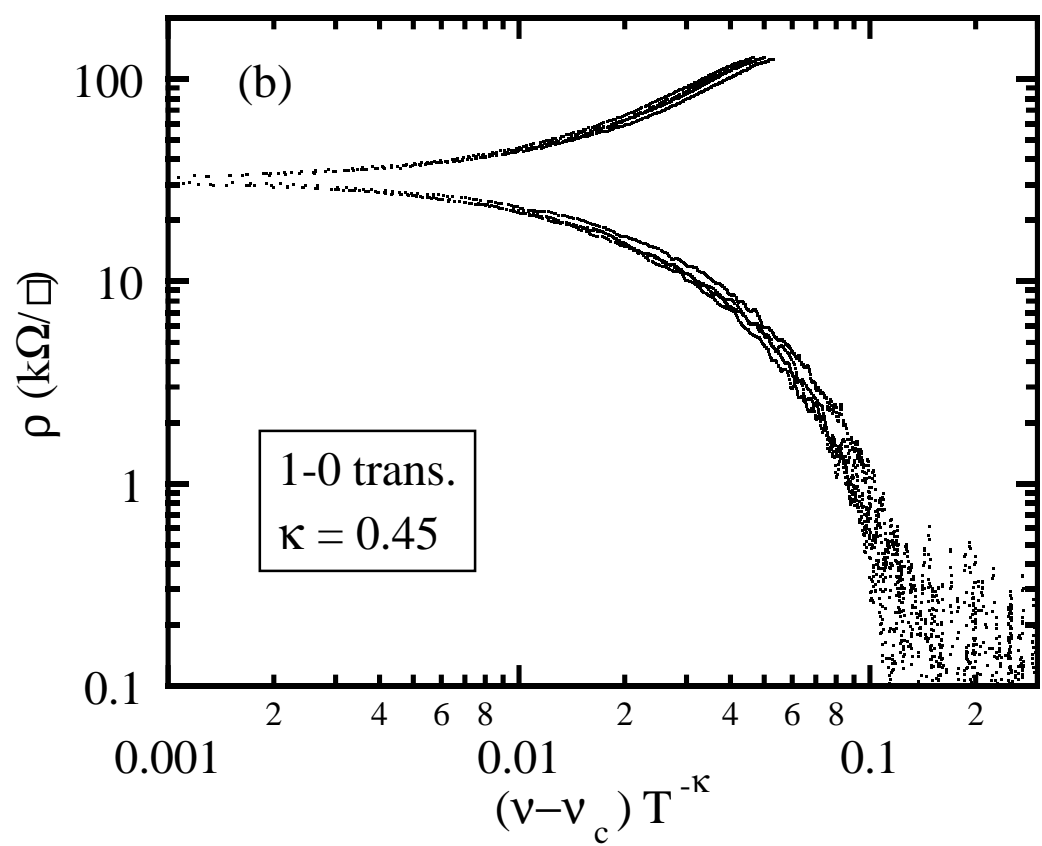
\end{document}